\documentclass[12pt]{JHEP3}
\usepackage{amsmath}

\title{Consistent modified gravity: dark energy,
 acceleration and the absence
of cosmic doomsday}
\author{M.C.B. Abdalla \\
Inst. Fisica Teorica, Universidade Estadual Paulista,
Sao Paulo, Brazil \\
email: mabdalla@ift.unesp.br}
\author{Shin'ichi Nojiri \\
Department of Applied Physics,
National Defence Academy,
Hashirimizu Yokosuka 239-8686, Japan \\
email: nojiri@nda.ac.jp, snojiri@yukawa.kyoto-u.ac.jp}
\author{Sergei D.Odintsov\thanks{Also at TSPU, Tomsk, Russia and at IFT, UNESP,
Sao Paulo (temporary)} \\
Instituci\`o Catalana de Recerca i Estudis
Avan\c{c}ats (ICREA)  and Institut d'Estudis Espacials de Catalunya (IEEC),
Edifici Nexus, Gran Capit\`a 2-4, 08034 Barcelona, Spain \\
email: odintsov@ieec.fcr.es, odintsov@ieec.uab.es}
\abstract{
We discuss the modified gravity which includes negative and positive powers
 of the curvature and which provides
the gravitational dark energy. It is shown that in GR plus the term  containing
negative power
of the curvature the cosmic speed-up may be achieved, while
the effective phantom phase (with $w$
less than $-1$) follows when such term  contains the fractional positive
power of the curvature. The minimal
coupling with matter makes the situation more interesting: even
$1/R$ theory coupled with the usual ideal fluid
may describe the (effective phantom) dark energy. The account of $R^2$ term
(consistent modified gravity)  may help to escape of cosmic doomsday.
}

\begin{document}

\tolerance=5000

\def\pp{{\, \mid \hskip -1.5mm =}}
\def\cL{{\cal L}}
\def\be{\begin{equation}}
\def\ee{\end{equation}}
\def\bea{\begin{eqnarray}}
\def\eea{\end{eqnarray}}
\def\beq{\begin{eqnarray}}
\def\eeq{\end{eqnarray}}
\def\tr{{\rm tr}\, }
\def\nn{\nonumber \\}
\def\e{{\rm e}}

\section{Introduction}

The interpretation of the very recent observational data indicates that current universe is flat and is
in accelerating expansion which started about five billion years ago. The important characteristic of the
(accelerated) expansion is so-called ``equation of state'' parameter $w$ (for dominant energy density
contribution) which is the relation between the pressure and the energy density of the universe.
It is estimated that about 70 percent of universe energy density is composed of some mysterious effective
fluid (dark energy which rules the universe) with negative pressure and $w$ being close to $-1$. Despite
the number of efforts we are still far away from the theoretical understanding of dark energy and its origin.

The very promising approach to dark energy is related with the (phenomenological) modifications of Einstein
gravity in such a way, that it would give the gravitational alternative to dark energy.
Conceptually, it looks very attractive as then the presence of dark energy is only the consequence of
the universe expansion. One such model containing $1/R$ term (which may originate from M-theory \cite{SN})
was proposed in ref.\cite{CDTT} as gravitational alternative for dark energy. Modified gravity describes
the accelerated expansion but it contains number of instabilities \cite{chiba}.
Nevertheless, further modification of the model by $R^2$-term \cite{SNPRD} or $\ln R$-term \cite{NOgrg}
(see also \cite{peng}) leads to consistent modified gravity which may pass solar system tests and is free of the
instabilities.\footnote{It is interesting that Palatini formulation of modified gravity leads to physically
different theory (for very recent discussion and list of refs., see \cite{palatini}) which seems to be
free of (some) instabilities too.} One may consider other generalizations like ones including the positive
(negative) powers of Ricci tensor squared \cite{mauro}, coupling of $f(R)$ theory with scalar
\cite{sergio,dominance}, multidimensional $1/R$ theory \cite{sasha} or even more extravagant
(non-symmetric) gravity (for recent discussion, see \cite{moffat}).
Definitely, various predictions of consistent modified gravity should be tested. In its own turn, these tests
may suggest further modifications giving true description of an observable universe.

The present work is devoted to further study of the properties of modified gravity which contains positive
and negative powers of the curvature.
It is demonstrated that (effective phantom) dark energy cosmological solutions for the model naturally contain
the finite-time, sudden singularity in the future (section 2). However, the consistent modified gravity with
negative (or positive, fractional) powers of the curvature as well as with $R^2$ term has more stable
future history. The cosmic doomsday does not occur there, rather the universe ends up in deSitter phase.
In the third section it is considered the coupling of modified gravity with the usual matter. The very interesting
observation shows that in such unified framework it is easier
to realize various types of effectively
dark energy universe.
In summary, some outlook is given.

\section{Sudden future singularity in modified gravity}

Let us start from the general model of gravity depending only on curvature:
\be
\label{XXX7}
S={1 \over \kappa^2}\int d^4 x \sqrt{-g} f(R)\ .
\ee
Here $f(R)$ can be an arbitrary function. 
It has been known \cite{wands} that such modified gravity theories can be rewritten into 
scalar-tensor 
 form via the conformal transformation. Depending on the form of the function f the 
scalar-tensor theory may contain ghost-like term (with negative kinetic energy).
It is quite remarkable that $f(R)$ gravity of special form as we show below 
from this big class of theories may find the interesting applications as candidate 
to describe
 dark energy universe and its acceleration. In a sense, it is return of somehow forgotten 
generalized gravity motivated by recent astrophysical data. 

By introducing the auxilliary fields $A$, $B$, one may rewrite the action (\ref{XXX7})  as
$S={1 \over \kappa^2}\int d^4 x \sqrt{-g} \left\{B\left(R-A\right) + f(A)\right\}$.
Using equation of motion and expressing $B$ in terms of $A$,  one arrives at the Jordan frame action.
Using the conformal transformation $g_{\mu\nu}\to \e^\sigma g_{\mu\nu}$ with $\sigma
= -\ln f'(A)$ (see also\cite{cotsakis}),
we obtain the Einstein frame action\cite{SNPRD}:
\bea
\label{XXX11}
&& S_E
={1 \over \kappa^2}\int d^4 x \sqrt{-g} \left( R - {3 \over 2}g^{\rho\sigma}
\partial_\rho \sigma \partial_\sigma \sigma - V(\sigma)\right)\ , \\
\label{XXX11b}
&& V(\sigma)= \e^\sigma g\left(\e^{-\sigma}\right) - \e^{2\sigma} f\left(g\left(\e^{-\sigma}
\right)\right) =  {A \over f'(A)} - {f(A) \over f'(A)^2}\ .
\eea
Note that two such theories in these frames are mathematically equivalent.
However, physics seems to be different. For instance, in Einstein frame
the matter does not freely fall along the geodesics which is well-established fact.

As an interesting example the action of (large distances) modified
gravity may be taken in the following form
\be
\label{BRX1}
S={1 \over \kappa^2}\int d^4 x \sqrt{-g} \left(R - \gamma R^{-n}\right)\ .
\ee
Here $\gamma$ is  (an extremely small) coupling constant and $n$ is some
number. Since the function $f(A)$ and the scalar field $\sigma$ are
\be
\label{XXX13}
f(A)=A -\gamma A^{-n}\ ,\quad
\sigma=-\ln \left(1 + n\gamma A^{-n-1}\right)
\ee
 the potential is given by
\be
\label{BRR2}
V=\frac{\gamma (n+1)A^{-n}}{\left(1 + n\gamma A^{-n-1}\right)^2}
= \gamma (n+1)\left(\frac{\e^{-\sigma} - 1}{n\gamma}\right)^{\frac{n}{n+1}}\e^{2\sigma}\ .
\ee
When curvature $(\sim A)$ is small and $n>-1$ and $n\neq 0$, the potential behaves as an exponential function
$V\sim \frac{1}{n\gamma}\left(1 + {1 \over n}\right) A^{n+2}
\sim \left(1 + {1 \over n}\right) \left(  n \gamma \right)^{1 \over n+1} \e^{{n+2 \over n+1}\sigma}$.
On the other side, when the curvature is large, it follows
$V \sim \gamma (n+1)A^{-n}\sim \frac{n+1}{n}\left( n \gamma \right)^\frac{1}{n+1}
\left( -\sigma \right)^{\frac{n}{n+1}}$.
Since
$V'(A)=-\frac{\gamma n(n+1)A^{-n-1}\left\{ 1 - (n+2)\gamma  A^{-n-1} \right\}}
{\left( 1 + n\gamma A^{-n-1}\right)^3}$,
$V(A)$ has only one extremum
and the extremum is given at $A=\left\{(n+2)\gamma \right\}^{\frac{1}{n+1}}$.

The FRW universe metric in the Einstein frame is chosen as
$ds_E^2= -dt_E^2 + a_E^2(t_E) \sum_{i=1}^3\left(dx^i\right)^2$.
If the curvature is small, the solution of  equation of motion is $a_E \sim t_E^\frac{3(n+1)^2}{(n+2)^2}$.
The FRW universe metric in Jordan frame is $ds^2= -dt^2 + a^2(t) \sum_{i=1}^3\left(dx^i\right)^2$,
where the variables in the Einstein frame and in the physical Jordan frame
are related with each other by $t= \int \e^{\frac{\sigma}{2}}dt_E$, $a=\e^{\frac{\sigma}{2}}a_E$.
This gives $t\sim t_E^{1 \over n+2}$ and
\be
\label{XXX14b}
a\sim t^{(n+1)(2n+1) \over n+2}\ ,\quad w=-{6n^2 + 7n - 1 \over 3(n+1)(2n+1)}\ .
\ee
The first important consequence of above Eq.(\ref{XXX14b}) is that there is possibility of accelerated
expansion for some choices of $n$ (effective quintessence). In fact if $n>\frac{-1+\sqrt{3}}{2}$ or
$-1<n<-\frac{1}{2}$, we find $w<-\frac{1}{3}$ and $\frac{d^2 a}{dt^2}>0$.\footnote{
In the solution (\ref{XXX14b}), the Hubble parameter $H\equiv\frac{1}{a}\frac{da}{dt}$ has the form of
$H=\frac{h_0}{t}$, where $h_0$ is a constant of the unity order. There is an ambiguity how to choose $t$
in  $H$.  One natural choice is to take $t$ to be of the order of the age $T$ of the
present universe. Since $T\sim 1.37\times 10^{10}$ years $\sim \left(10^{-33}\ {\rm eV}\right)^{-1}$,
we find $H\sim 10^{-33}\ {\rm eV}$, which is consistent with the observed value of $H$ in the
present universe: $H_{\rm observed}\sim 70\,{\rm km\,s}^{-1}{\rm Mpc}^{-1} \sim 10^{-33}\,{\rm eV}$.
Then even if the original Lagrangian theory does not contain small parameter of the order of the
Hubble parameter, such a small scale is naturally  induced from the age of the universe in the power law expansion
 (\ref{XXX14b}).}
In other words, modified gravity presents the gravitational alternative for dark energy with the possibility
of cosmic speed-up. The corresponding analysis was given in detail in ref.\cite{NOgrg}.

When $-1<n<-\frac{1}{2}$, it follows $w<-1$.
If $w<-1$, the universe is shrinking in the expression of $a$ (\ref{XXX14b}). If we replace the
direction of time by changing $t$ by $-t$, the universe is expanding but  $t$ should be considered
to be negative
so that the scale factor $a$ should be real. Then there appears a singularity at $t=0$,
where the scale factor $a$ diverges as $a \sim \left(  - t \right)^\frac{2}{3(w+1)}$.
One may shift the origin of the time by further changing $-t$ with $t_s - t$. Hence, in the present universe,
$t$ should be less than $t_s$ and it looks to appear the singularity at $t=t_s$:
$a \sim \left(t_s  - t \right)^\frac{2}{3(w+1)}$.
The future singularity may be called sudden or Big Rip singularity.
We should note, however, the expression of $w$ in (\ref{XXX14b}) could be correct only when curvature is small,
as in the present universe.
When $t$ goes to $t_s$, the curvatures become large as $\left(t_s - t\right)^{-2}$ and Eq.(\ref{XXX14b})
becomes invalid. As the curvature becomes large, the first Einstein-Hilbert term in (\ref{BRX1})
 dominates.
Then as will be shown later, the singularity is moderated and if we include usual matter,
the singularity will not appear eventually.
Since $w+1=\frac{2(n+2)}{3(n+1)(2n+1)}$, it follows $w<-1$ when $-1<n<-\frac{1}{2}$ (the effective phantom
phase for fractional positive power of the curvature). Here it is assumed $n>-1$, so that the Einstein term
dominates in (\ref{XXX13}) when the curvature is small.
We should also note when $n>0$, as $\frac{2(n+2)}{3(n+1)(2n+1)}>0$, $w$ is always greater than $-1$.
Then the theory with negative power of the curvature like $\frac{1}{R}$-gravity does not produce effective
phantom although it may produce effective quintessence.

The results (\ref{XXX14b}) are valid when the curvature is small but near the Big Rip singularity,
the curvature becomes large and (\ref{XXX14b}) is not valid. The qualitative behavior when the curvature
is large can be found from the potential.

The qualitative behavior of the potential when $-1<n<-\frac{1}{2}$ and $\gamma>0$ is given
in Figure \ref{Fig1}a). In order that $\sigma$ is real, however, Eq.(\ref{XXX13}) tells
$R\sim A> \left( -n\gamma \right)^\frac{1}{n+1}$.
Then the curvature cannot be small and the expressions (\ref{XXX14b}) are not valid.
When $A< \left( -n\gamma \right)^\frac{1}{n+1}$, instead of (\ref{XXX13}), one may define
$\sigma=-\ln \left(-1 - n\gamma A^{-n-1}\right)$. As shown in \cite{NOgrg}, however, the anti-gravity
appears in this case and instead of (\ref{XXX11}), we obtain
$S_E={1 \over \kappa^2}\int d^4 x \sqrt{-g} \left( - R + {3 \over 2}g^{\rho\sigma}
\partial_\rho \sigma \partial_\sigma \sigma - \tilde V(\sigma)\right)$ with potential
$\tilde V(\sigma)= -\e^\sigma g\left(\e^{-\sigma}\right) - \e^{2\sigma} f\left(g\left(\e^{-\sigma}
\right)\right)$.
Hence, when $-1<n<-\frac{1}{2}$ and $\gamma>0$, the region $A< \left(-n\gamma \right)^\frac{1}{n+1}$
is not  physical. Then we should assume $\gamma<0$.
In case $-1<n<-\frac{1}{2}$ and $\gamma<0$, there does not appear the extremum in the potential
if curvature is positive. The qualitative behavior of the potential is given in
Figure \ref{Fig1}b). The Eq.(\ref{BRR2}) shows that the potential is negative and is unbounded below
when $A$ is large.

\FIGURE 
{
{\small
\parbox[c]{6cm}{
\unitlength=0.3mm
\begin{picture}(160,100)
\thicklines
\put(20,20){\vector(1,0){120}}
\put(20,20){\vector(0,1){60}}
\qbezier[300](20,20)(49,20)(49,80)
\qbezier[300](51,80)(51,30)(80,30)
\qbezier[300](80,30)(110,30)(140,70)
\put(20,82){\makebox(0,0)[b]{$V(A)$}}
\put(142,18){$A=R$}
\put(50,18){\makebox(0,0)[t]{$C_1$}}
\put(80,18){\makebox(0,0)[t]{$C_2$}}

\put(80,0){\makebox(0,0)[t]{a)}}

\thinlines
\qbezier[60](50,20)(50,50)(50,80)
\qbezier[10](80,20)(80,25)(80,30)
\end{picture}
}
\parbox[c]{6cm}{
\unitlength=0.3mm
\begin{picture}(160,100)
\thicklines
\put(20,80){\vector(1,0){120}}
\put(20,80){\vector(0,-1){60}}
\qbezier[800](20,80)(60,80)(140,20)
\put(20,18){\makebox(0,0)[t]{$V(A)$}}
\put(142,78){$A=R$}

\put(80,0){\makebox(0,0)[t]{b)}}

\end{picture}
}}
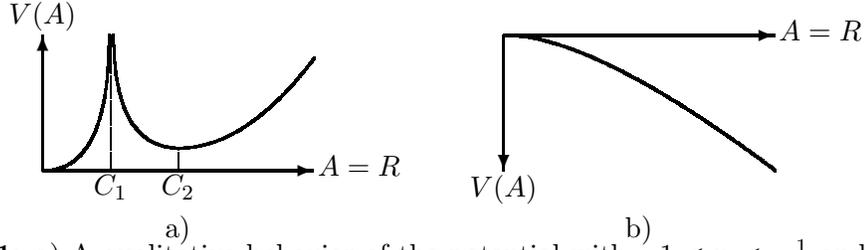
\caption{a) A qualitative behavior of the potential with $-1<n<-\frac{1}{2}$ and $\gamma>0$.
Here $C_1\equiv \left( -n\gamma \right)^\frac{1}{n+1}$ and
$C_2\equiv \left( -\left(n+2\right)\gamma \right)^\frac{1}{n+1}$
In the region with $A<\left( -n\gamma \right)^\frac{1}{n+1}$, there
appears the anti-gravity phase.
b). A qualitative behavior of the potential with $-1<n<-\frac{1}{2}$ and
$\gamma<0$.
The potential is monotonically decreasing.
\label{Fig1}}
}

In order to consider the region where the curvature is large for the case $-1<n<-\frac{1}{2}$ and $\gamma<0$,
the following potential is taken
\be
\label{BRR6}
V = -V_0 \left( -\sigma \right)^\alpha \ ,\quad
V_0 \equiv - \frac{n+1}{n}\left( n \gamma \right)^\frac{1}{n+1}>0\ ,\quad \alpha \equiv \frac{n}{n+1}<-1\ .
\ee
The   $\sigma$-equation of motion and the FRW equation in the Einstein frame are
$0=- {3} \left(\frac{d^2 \sigma}{dt_E^2} + 3H \frac{d\sigma}{dt_E}\right) - V'(\sigma)$ and
${6 \over \kappa^2}H_E^2={3 \over 2}\left(\frac{d\sigma}{dt_E}\right)^2 + V(\sigma)$.
A consistent solution is given by
\be
\label{BRR7}
H_E=0,\ \sigma= - \sigma_0 \left(t_{sE} - t_E\right)^\beta,\
\sigma_0\equiv \left(\frac{2V_0}{3\beta^2}\right)^{\frac{1}{2-\alpha}},\
\beta\equiv \frac{2}{2-\alpha}=\frac{2(n+1)}{n+2}\ .
\ee
Here $t_{sE}$ is a constant of  integration.
Since $-1<n<-\frac{1}{2}$, it follows
$0<\beta<\frac{2}{3}$ .
The spacetime in the Einstein frame is flat and the scale factor $a_E=a_0$ is a constant which
is not the case in the physical Jordan frame, where
 they are finite even at $t_E=t_{sE}$ although there is a cut singularity there.
The cut singularity, however, makes the physical curvature divergent.
In fact, when $t_E\sim t_{sE}$, $t$ is given by
$t=t_s + t_E - t_{sE}$ with a constant of  the integration $t_s$.
In the Jordan frame, the scale factor $a$ is given by
$a(t) \sim a_0 \e^{-\frac{\sigma_0}{2}\left(t_s - t\right)^\beta}$
and the Hubble parameter $H\equiv \frac{1}{a}\frac{da}{dt}$ behaves as
$H=\frac{1}{2}\frac{d\sigma}{dt}= \frac{1}{2}\frac{d\sigma}{dt_E}\frac{dt_E}{dt}
= \frac{\beta\sigma_0}{2}\left(t_{sE} - t_E\right)^{\beta -1} \e^{-\frac{\sigma}{2}}
\sim \frac{\beta\sigma_0 }{2}\left(t_s - t\right)^{\beta -1}$.
Since $-1<\beta - 1 <-\frac{1}{3}<0$, $H$ diverges at
$t=t_s$ as well as the scalar curvature $R=6\frac{dH}{dt} + 12 H^2$.
In case of the Big Rip singularity, $H$ behaves as $H\sim \frac{h_0}{t_s - t}$.
Hence, the behavior of $H$ is moderated. In case of the Barrow model \cite{Barrow}
(see also \cite{Barrow1,varun}), $H$ behaves as $H\sim h_0 + h_1\left(t_s - t\right)^\alpha$ with constant $h_0$
and $h_1$. As $0<\alpha<1$, the behavior of $H$ here is more singular than that  in \cite{Barrow}.
The singularity in the present case is moderated and $a$ does not diverge.
With  the usual matter inclusion such a singularity will not appear.
It is interesting to note that for the model \cite{Barrow},
 the singularity is moderated (or even disappears)
if the quantum corrections are taken into account \cite{BarrowQC}. Note also that usual, infinite-time singularity
is still possible even in consistent modified gravity (for earlier, related discussion of infinite time, future
singularity in $f(R)$ gravity, see \cite{Barrow2}).

In accord with the proposal of ref.\cite{SNPRD}
one may add a term proportional to $R^2$ to the action (\ref{BRX1})
\be
\label{BRXXX1}
S={1 \over \kappa^2}\int d^4 x \sqrt{-g} \left(R - \gamma R^{-n} + \eta R^2\right)\ .
\ee
In this situation the unified theory permits early time inflation as well as late time acceleration and does not
contain the number of instabilities.
The function $f(A)$  (\ref{XXX13}) is modified to be $f(A)=A -\gamma A^{-n} + \eta A^2$.
We now assume $-1<n<-\frac{1}{2}$, $\gamma<0$, and $\eta>0$.
Then  $\sigma = - \ln \left(1 + n\gamma A^{-n-1} + 2\eta A\right)$.
Although $1 + n\gamma A^{-n-1} + 2\eta A>0$, since
$\frac{d\sigma}{dA}=-\frac{f''(A)}{f'(A)}=-\frac{-n(n+1)\gamma A^{-n-2} + 2\eta}{1 + n\gamma A^{-n-1} + 2\eta A}$,
there is a branch point, where $\frac{d\sigma}{dA}=0$ or $f''(A)=0$, at
$A=A_0\equiv \left\{\frac{n(n+1)\gamma}{2\eta}\right\}^{\frac{1}{n+2}}$ or
$\sigma=\sigma_0 \equiv - \ln \left( 1 + (n+2)\left(\frac{2\eta}{n+1}\right)^\frac{n+1}{n+2}
(n\gamma)^\frac{1}{n+2} \right)$.
The potential $V(A)$ has the following form:
\be
\label{BRR14}
V(A)=\frac{(n+1)\gamma A^{-n} + \eta A^2}{\left(1 + n\gamma A^{-n-1} + 2\eta A\right)^2}\ .
\ee
The behavior when $A$ is small is not changed from the case in (\ref{BRX1}). On the other hand, when $A$ is large, $V(A)$ goes
to a constant: $V(A)\to \frac{1}{4\eta}$.
The potential $V(A)$ vanishes at $A=A_1\equiv \left\{\frac{-(n+1)\gamma}{\eta}\right\}^{\frac{1}{n+2}}$.
Note $A_0<A_1$ since $0<\frac{n\gamma}{2}<-\gamma$.
$V'(A)$ has an extremum at $A=A_0$ since $V'(A)=\frac{\left\{-n(n+1)\gamma A^{-n-2} + 2\eta\right\}
A\left\{1-(n+2)\gamma A^{-n-1}\right\}}{\left(1 + n\gamma A^{-n-1} + 2\eta A\right)^3}$.
Since there is a branch point at $\sigma=\sigma_0$,
if we start from the small curvature,
the growth of the curvature stops at $R=A_0$, where $\sigma=\sigma_0$. In fact, at the branch point,
where $f''(A)=0$, the mass $m_\sigma$ of $\sigma$ becomes infinite since
$m_\sigma \propto \frac{d^2 V}{d\sigma^2}
= \frac{f'(A)}{f''(A)}\frac{d}{dA}\left(\frac{f'(A)}{f''(A)}\frac{dV(A)}{dA}\right)
= -\frac{3}{f''(A)} + \frac{A}{f'(A)} + \frac{2f(A)}{f'(A)^2} \to + \infty$.
Note also $f''(0)<0$ when $A<A_0$.
Then the growth of $\sigma$ is finished at $\sigma=\sigma_0$.
Hence, adding $R^2$ term, there does not occur cosmic doomsday but the
universe ends up in deSitter phase. The scale factor $a$
is given by $a(t)\sim a_0\e^{t \sqrt{\frac{A_0}{12}}}$ with a constant $a_0$.
Note that the quantities in the Einstein frame are different from those in the Jordan frame by almost
constant factor as $a_E\sim \e^\frac{\sigma_0}{2}$ or $dt_E\sim \e^\frac{\sigma_0}{2} dt$.
This supports our point of view that cosmic doomsday in such theory does not occur because
after the late time acceleration the universe starts new inflationary era.
It is also important to stress that as the mass of $\sigma$ becomes very large, there is no problem about
the equivalence principle since $\sigma$ cannot mediate the force.
In other words, unlike to BD theory, such a model may pass the solar system test provided by VLBI experiment.

\section{Modified gravity coupled with matter}

It is very interesting that modified gravity which can be made consistent one \cite{SNPRD} may help
in the resolution of dark energy problem in various ways as it suggests gravitational alternative for dark energy.
In particular, as we will show below it may give phantom  dark energy without necessity to introduce
the (negative kinetic energy) phantom scalar theory. In fact, the matter is taken to be the usual ideal fluid.

We now consider the system of the modified gravity coupled with matter:
\be
\label{M1}
S = \int d^4 x \sqrt{-g}\left\{ f(R) + L_m\right\}\ .
\ee
Here $f(R)$ is an adequate function of the scalar curvature and $L_m$ is a matter Lagrangian.
Then the equation of the motion is given by
$0=\frac{1}{2}g_{\mu\nu} f(R) - R_{\mu\nu}f'(R) - \nabla_\mu \nabla_\nu f'(R)
 - g_{\mu\nu}\nabla^2 f'(R) + \frac{1}{2} T_{\mu\nu}$.
Again, the FRW spacetime is considered.
The ideal fluid is taken as the matter with the constant $w$: $p=w\rho$.
Then from the energy conservation law it follows
$\rho = \rho_0 a^{-3(1+w)}$ .
In a some limit, strong cuvature or weak one, $f(R)$ may behave as
$f(R)\sim f_0 R^\alpha$ ,
with constant $f_0$ and $\alpha$.
 An exact solution of the equation of motion
is found to be
\bea
\label{M8}
&& a=a_0 t^{h_0} \ ,\quad h_0\equiv \frac{2\alpha}{3(1+w)} \ ,\nn
&& a_0\equiv \left[-\frac{6f_0h_0}{\rho_0}\left(-6h_0 + 12 h_0^2\right)^{\alpha-1}
\left\{\left(1-2\alpha\right)\left(1-\alpha\right) - (2-\alpha)h_0\right\}\right]^{-\frac{1}{3(1+w)}}\ .
\eea
When $\alpha=1$, the result $h_0 = \frac{2}{3(1+w)}$ in the Einstein gravity is reproduced.
Note that stability issue should be carefully investigated here.
However, even if the solution is instable the decay time could be very big due to the fact that coupling constant
of modified gravity term is very small. From another side, when finite-time future singularity occurs it may
be resolved by the account of quantum effects \cite{BarrowQC,NOgrg}.

The effective $w_{\rm eff}$ may be defined by $h_0=\frac{2}{3\left(1+w_{\rm eff}\right)}$.
By using (\ref{M8}), one finds
\be
\label{M9}
w_{\rm eff}=-1 + \frac{1+w}{\alpha}\ .
\ee
Hence, if $w$ is greater than $-1$ (effective quintessence or even usual ideal fluid with positive $w$),
when $\alpha$ is negative, we obtain the effective phantom phase where $w_{\rm eff}$ is less than $-1$.
This is different from the case of pure modified gravity.
Moreover, when $\alpha>w+1$ (it can be even positive),  $w_{\rm eff}$ could be negative (for negative $w$).
Hence, it follows that modified gravity minimally coupled with usual (or quintessence) matter may reproduce
quintessence (or phantom) evolution phase for dark energy universe in an easier way than without such coupling.

 One may now take $f(R)$ as in (\ref{BRXXX1}).
When the cuvature is small, the second term becomes dominant and one may identify
$f_0=-\frac{\gamma}{\kappa^2}$ and $\alpha=-n$. Then from (\ref{M9}),
it follows
$w_{\rm eff}=-1 - \frac{1+w}{n}$ .
Hence, if $n>0$, we have an effective phantom even if $w>-1$. Usually
the phantom generates
the Big Rip singularity. However, near the Big Rip singularity, the curvature becomes large and the
last term  becomes dominant. In this case $\alpha=2$ and
$w_{\rm eff}=\frac{-1+w}{2}$ .
Then if $w>-1$, it follows $w_{\rm eff}>-1$, which prevents
the Big Rip singularity (makes phantom phase transient) as is described
in the previous section.
To conclude, it looks quite promising that modest modification of General Relativity coupled to ideal fluid
matter leads to effective dark energy universe in the very natural way.

\section{Discussion}

In summary, the gravity theory with negative (like $1/R$) and positive (quadratic) powers of scalar curvature
shows the number of features which are desireable to explain the accelerating dark energy universe:

\noindent
1. It passes solar system tests (VLBI experiment). Thanks to higher derivative $R^2$ term, the gravitationally
bound objects (like Sun or Earth) are stable as well as in GR. Of course, some fine-tuning of coupling constants
is necessary as is shown in \cite{SNPRD}. (It was mentioned already in ref.\cite{CDTT} that the coefficient of
$1/R$-term should be extremely small.)

\noindent
2. Newtonian limit is recovered just at the above values for coupling constants.

\noindent
3. The gravitational dark energy
 dominance is explained simply by the universe expansion. Moreover, when modified gravity is coupled with
usual matter it is easier (less deviations from GR) to get the effective (phantom or quintessence)
dark energy universe regime.

\noindent
4. The presence of $R^2$ term in consistent modified gravity
 prevents the development of the cosmic doomsday.

Thus, the consistent modified gravity remains to be  viable candidate for the explanation of dark energy
as the gravitational phenomenon. Nevertheless, only future, more precise astrophysical/gravitational data
will prove if it is the time for new gravitational physics to enter the game.

\section*{Acknoweledgments}

We thank J.D. Barrow for helpful discussions.
This research has been supported in part by the Monbusho
 of Japan under grant n.13135208 (S.N.), by grant n.302019/2003-0
of CNPq, Brazil (M.C.B.A.),
by grant 2003/09935-0 of FAPESP, Brazil (S.D.O.)
 and by project BFM2003-00620, Spain (S.D.O.).

\end{document}